\documentclass[%
 reprint,
 showpacs,
 amsmath,amssymb,
 aps,
]{revtex4-1}

\usepackage{epsfig,graphicx}
\usepackage{dcolumn}
\usepackage{bm}

\begin{document}

\title{\bf Search for 2K(2$\nu$)-capture of $^{78}$Kr}


\author{
Ju.M.~Gavriljuk$^a$,
A.M.~Gangapshev$^{a}$, 
V.V.~Kazalov$^{a}$,
V.V.~Kuzminov$^{a}$, \\
S.I.~Panasenko$^{b}$,
S.S.~Ratkevich$^{b}$,
D.A.~Zhantudueva$^{a}$,
S.P. Yakimenko$^{a}$
\\
$^a$\small{\em Baksan Neutrino Observatory INR RAS} \\
\small{\em pos.~Neitrino, Elbrus raion, 361609 Kabardino-Balkaria, Russia} \\
$^b$\small{\em Karazin Kharkiv National University, Ukraine} \\
\small{\em Svobody Sq. 4, 61022, Kharkiv. Ukraine}
}
\date{May 26, 2011}

\begin{abstract}
Results of a search for $^{78}$Kr double $K$-capture at the second stage of the experiment
with high-pressure copper proportional counters are presented.
The method is based on comparison of spectra measured with natural and enriched krypton.
The total exposure of the low background measurements is for $^{78}$Kr 152 g$\times$yr and for $^{nat}$Kr 106 g$\times$yr.
An excess of events was observed in the analysis of $^{78}$Kr selected data collected during 2008-2010 years.
This excess could correspond to a double $K$-capture of $^{78}$Kr with the half-life of
$\texttt{T}_{1/2}(2\texttt{K},2\nu+0\nu)=[1.4^{+2.2}_{-0.7}]\cdot 10^{22}$ yr (90\% C.L.)
\end{abstract}


\maketitle

\section{Introduction}
The low background laboratory of the BNO INR RAS carries out researches
into rare processes and decays.  $2K(2\nu)$ capture in $^{78}$Kr is a
rare process that has not been observed yet. Theoretical calculations
for $2K(2\nu)$ within the frames of different models (QRPA, MSM and SU(4)$_{\sigma\tau}$)
predict the following half-life periods with respect to $2K$-capture:
$3.7 \cdot 10^{21}$ yr \cite{r1},  $4.7 \cdot 10^{22}$ yr  \cite{r2},
and $7.9 \cdot 10^{23}$ yr \cite{r3}. The values from works \cite{r2} and \cite{r3}
were obtained by taking the fraction of $2K(2\nu)$-capture events
in $^{78}$Kr with respect to the total number of $2e(2\nu)$-capture
events to be 78.6\% \cite{r4}. By comparing experimental and theoretical
results one can see that sensitivity of measurements has already passed beyond the
lower margin of theoretical predictions. This fact stimulates the research to
test the approach used in model \cite{r2}.

\section{Basic assumptions}
When two electrons are captured from the \emph{K}-shell in $^{78}$Kr, a daughter
atom of $^{78}$Se$^{**}$ is formed with two vacancies in the \emph{K}-shell.
The technique to search for this reaction is based on the assumption
that the energies of characteristic photons and the probability that
they will be emitted when a double vacancy is filled are the same as the
sum of respective values when single vacancies of the \emph{K}-shell in
two singly ionized Se$^*$ atoms are filled. In such a case, the total
measured energy is $2K_{ab}= 25.3$ keV, where ${K_{ab}}$
is the binding energy of a $K$ electron in a Se atom (12.65 keV). The
fluorescence yield upon filling a single vacancy in the \emph{K}-shell of
Se is 0.596. The energies and relative intensities of the
characteristic lines in the $K$ series are $K_{\alpha1}=11.22$ keV (100\%),
$K_{\alpha2}=11.18$ keV (52\%), $K_{\beta1}=12.49$ keV (21\%), and $K_{\beta2}=12.65$ keV (1\%) \cite{r5}.
There are three possible ways for de-excitation of a doubly ionized
\emph{K}-shell: 1) emission of Auger electrons only $(e_a, e_a)$;
2) emission of a single characteristic quantum and an Auger electron
$(K,e_a)$; and 3) emission of two characteristic quanta and
low-energy Auger electrons $(K,K,e_a)$, with probabilities of
$p_1 = 0.163$, $p_2 = 0.482$, and $p_3 = 0.355$, respectively. A characteristic
quantum can travel a long enough distance in a gas medium between the
points of its production and absorption. For example, 10\% of characteristic quanta with energies of 11.2 and 12.5 keV are absorbed in krypton at a pressure of 4.35 atm ($\rho = 0.0164$ g/cm$^3$) on a path of 1.83 and 2.42 mm long, respectively (the values of absorption factors are taken from \cite{r6} and \cite{Storm}). The paths of photoelectrons with the same energies are 0.37 and 0.44 mm, respectively. They produce almost pointwise charge clusters of primary ionization in the gas. In case of the event with the escape of two characteristic quanta absorbed in the working gas and a single Auger electron, the energy is distributed among three pointwise charge clusters. It is these three-point (or three-cluster) events possessing a unique set of features that were the subject of the search in \cite{Kaz09} and \cite{wav10}.

To register the process of $2K$-capture in
$^{78}$Kr a large proportional counter (LPC) with a casing of
M1-grade copper has been used.
The LPC is inside the low-background shield made of 18 cm copper + 15 cm lead + 8 cm borated polyethylene.
The installation is placed in one of the chambers of the underground laboratory of the Gallium Germanium Neutrino Telescope experiment at the Baksan Neutrino Observatory, INR RAS at a depth of 4700 m.w.e.
where cosmic ray flux is lowered by $\sim 10^7$ times to the level of $(3.03 \pm  0.10) \cdot 10^{-9}$ cm$^{-2}$s$^{-1}$ \cite{Gavrin91}.

The output signals from the LPC are registered by a digital oscilloscope. Description of a recorder system and methods of multipoint event identification and their processing are described in details in ref. \cite{Kaz09}, \cite{wav10} and \cite{PTE}.
Results of processing of the data measured in 2008-2010 years are presented in the present paper.

\section{Measurement results}

\begin{figure}[pt]
\includegraphics[width=8cm,angle=0.]{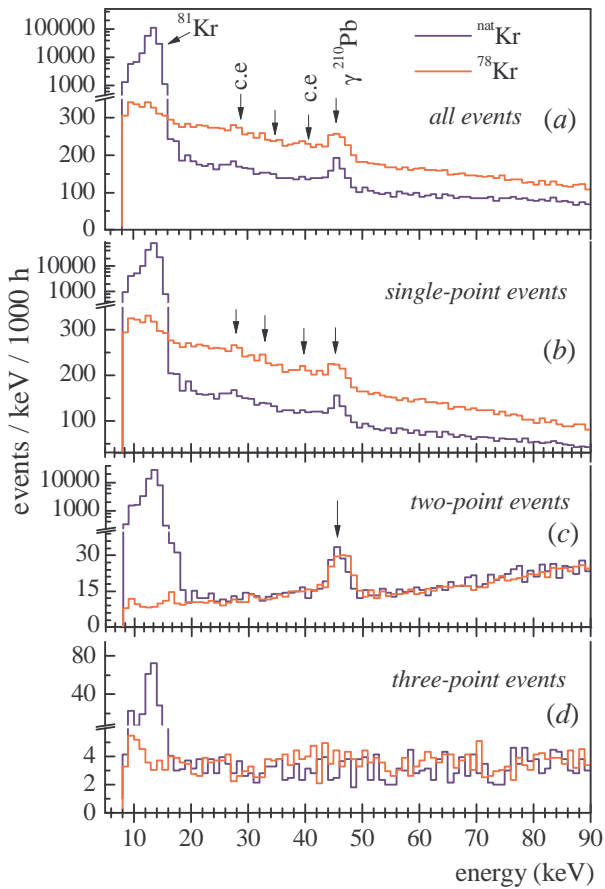}%
\caption{\label{f1}Spectrum (normalized for 1000 h) of the background of LPC filled with krypton enriched in $^{78}$Kr (orange curve) and with natural krypton (blue curve): (\emph{a}) - all events, (\emph{b}) - single-point events, (\emph{c}) - two-point events and (\emph{d}) - three-point events.}
\end{figure}

In Fig.\ref{f1}\emph{a} one can see total spectra, normalized for 1000 h, obtained for the background of LPC filled with krypton enriched in $^{78}$Kr (total collection time is 9457 h; orange curve), and with natural krypton (total collection time is 6243 h; blue curve).
Comparison of these spectra shows that the background of the counter filled with krypton enriched in $^{78}$Kr exceeds the background with natural krypton at energies above 18 keV.

The shape of difference spectrum in the range of $20\div100$ keV is described well enough by the model spectrum composed of $\beta$-spectra due to $^{85}$Kr and $^{14}$C decay in proportion 1:17. Background counting rates in this range are 56.3 h$^{-1}$ and 47.2 h$^{-1}$ for LPC filled with $^{78}$Kr and $^{nat}$Kr, respectively. Difference in $^{85}$Kr isotope activities in the samples is due to its different residual value achieved during the isotopic purification. $^{14}$C isotope comes into the working gas during the continuous usage of the counter, supposedly as a result of slow sublimation of organic molecules (ethyl alcohol and acetone were used to clean the details at the time of mounting the detector) from the surface of the body of the counter. This overactivity of $^{14}$C in the enriched krypton is due to the fact that this sample was measured first when the main portion of this organic vapor was there. Actually, all the events in the operating energy range produced by $\beta$-particles absorption can be considered as single-point ones. Except for those events where $\beta$-particles have lost partially their energy through generating bremsstrahlung photon.

At energies  $\leq 18$ keV, Fig.\ref{f1}, the counting rate of events in
spectrum of natural krypton is much higher than in spectrum of
krypton enriched in $^{78}$Kr. This is due to the presence, in the original
atmospheric krypton, of cosmogeneous radioactive isotope $^{81}$Kr.
Counting rate of events in this energy range is 220 h$^{-1}$,
which corresponds to the volume activity of $^{81}$Kr $(0.10\pm0.01)$
min$^{-1}l^{-1}$Kr.

At 46.5 keV, in spectra from the sample natural and enriched krypton Fig.~\ref{f1}\emph{a}, one can see the peak corresponding to the source
of $^{210}$Pb (T$_{1/2}=22.2$ yrs, $\beta$-decay, $E_\gamma=46.5$ keV, with a yield of 4.25\% per decay \cite{nudat2}).

Spectra of Fig.\ref{f1}\emph{d} have been used for further analysis of the three-point events.
Each event of these spectra is characterized by a set of energy deposits distributed over three point-like regions of the working volume of the counter.
Energy deposits are proportional to the Gaussian areas ($A_1$, $A_2$, $A_3$), which describe three pointwise components
of the composite current pulse of primary ionization electrons on the border of the gas amplification region.
The Gaussian numbering corresponds to the arrival time sequence of the components coming into the gas amplification region.
To simplify the selection of events with a given set of features the amplitudes of components for each event  are arranged in the
increased order $[(A_1, A_2, A_3) \rightarrow (q_0 \leq q_1 \leq q_2)]$.
In the sought-for events the minimal amplitude value (two groups of Auger-electrons from residual excitation with energies
of $E_1 \sim 1.5$ keV and  $E_2 \sim2.9$ keV)
will correspond, with superfluously exhibited resolution, to the condition ("\emph{C1}") 0.9 keV $\leq q_0 \leq 4.5$ keV.
To select useful events we introduce a parameter of ratio of average amplitude to maximum one.

\begin{figure}[pt]
\begin{center}
\includegraphics*[width=8cm]{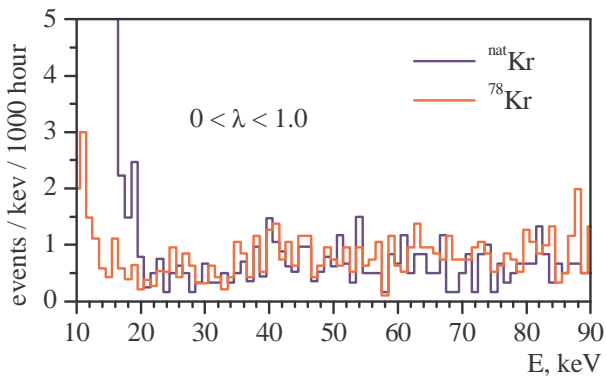}
\caption{\label{f2}The three-point spectra selected under the conditions "\emph{C1}" and "\emph{C2}".}
\end{center}
\end{figure}

\begin{figure}[pt]
\includegraphics*[width=8cm]{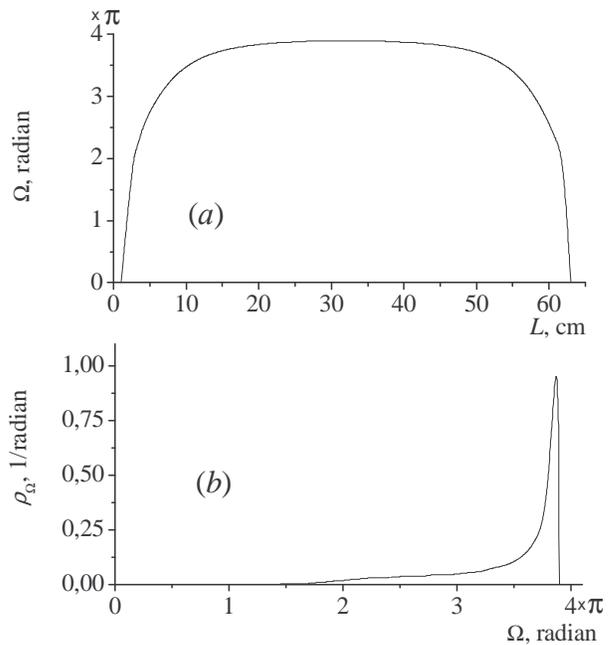}
\caption{\label{f3}\emph{a} - the solid angle through which the interior surface of the counter from different points of an anode wire is viewed. \emph{b} - the density of distribution of the solid angle for the points
uniformly distributed along the anode wire.}
\end{figure}

With an allowance for the resolution most part of useful events would have this condition ("\emph{C2}") fulfilled within
$1.0\geq q_1/q_2\geq0.65$.
The selection of events from spectra in Fig.~\ref{f1}\emph{d} corresponding to the conditions ("\emph{C1}"+"\emph{C2}") allows an additional decrease in the background of the expected peak energy region and gives a general representation of spectrum shape for wide range of energies. The selected spectra are plotted in Fig.\ref{f2}.
From the figure, we can see, that both spectra have a similar pick in the energy region of $\sim 26$ keV corresponding to the sought-for effect energy region. To find their nature the study of the distribution of the number of background events of various types along the anode wire has been performed. This distribution was plotted using the dependence of the relative amplitude of the first after-pulse on the distance between the point of origin of the main pulse and the middle point of the length of the anode wire. This distance defines the solid angle viewing the inner surface of the copper cathode cylinder. The solid angle through which the working surface of the cathode is viewed from the middle point of the anode wire and from the end points of the working length of the anode wire is $\sim 3.9\pi$ and $\sim2\pi$ , respectively, Fig.\ref{f3}\emph{a}.

The density of distribution of the solid angle for the points uniformly distributed along the anode wire are presented in Fig.\ref{f3}\emph{b}. There is a dependence of the relative number of photoelectrons, knocked out of the copper surface by the photons produced in the gas amplification of the primary ionization, on the value of the solid angle. First after-pulse is produced as a consequence of gas amplification of secondary photoelectrons generated on the cathode within the working length of the anode wire.

It is $\lambda$-parameter, equal to the ratio of first after-pulse M2 and pulse M1 amplitudes, $\lambda=100\times(M2/M1)$, accurate to the precision set by energy resolution of pulse and after-pulse, that determines the coordinate of the primary event with respect to the middle point of the anode wire.

\begin{figure}[pt]
\includegraphics*[width=8.0cm,angle=0.]{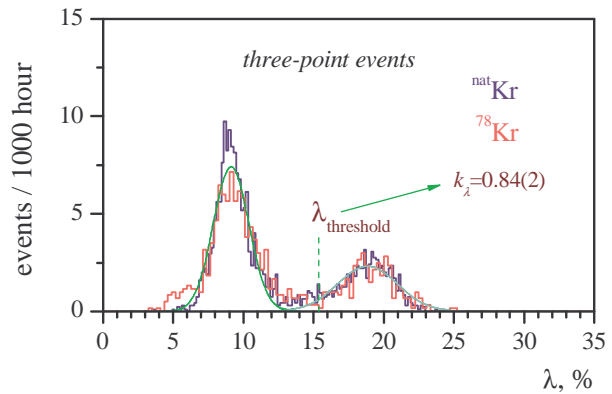}
\caption{\label{f4} $\lambda$-distributions of three-point events in LPC filled with krypton enriched in
$^{78}$Kr (orange curve), and with natural krypton (blue curve).}
\end{figure}

Fig.\ref{f4} demonstrates $\lambda$-distribution for the background of LPC filled with Kr enriched in $^{78}$Kr (orange curve) and with natural Kr (blue curve). One can see the excess of events in the histogram in the region with $\lambda$ taken from the ends of the anode wire.

These additional events at the ends of the working length of the anode wire are produced by particles coming out of the 'dead' near-end volumes of a working gas. In addition to proportionally amplified components a total pulse could be composed of components of larger energy deposits that occur in the 'dead' volume and are collected in the ionization mode at the end bulbs of the anode.

Specifically, the peak at $\sim26$ keV in spectra of Fig.\ref{f2} could well be formed by alpha-particles emitted from the surface of the cathode in the near-end region where the anode wire is thickened with copper tubules. Alpha-particle can ionize \emph{K}-shells of krypton atoms, and in this case the characteristic photons would have energy of 12.6 keV. These photons can penetrate into the working volume of LPC where gas amplification occurs. Registration of two photons gives two points. The third point is produced by ionization of alpha-particle collected in the ionization mode. By discarding pulses with   $\lambda \leq 15.5$, one can completely eliminate such events from 'three-point' spectra. The selection coefficient of useful event $k_\lambda$ remains high enough: $k_\lambda=0.84$.

\begin{figure}[pt]
\includegraphics*[width=8.0cm,angle=0.]{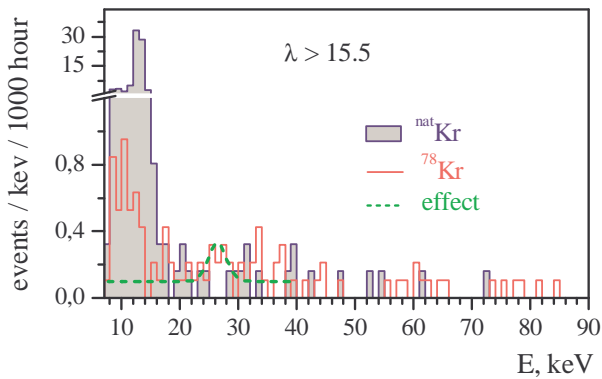}
\caption{\label{f5}The three-point spectra selected under the conditions,"\emph{C1}"+"\emph{C2}", and for $\lambda \geq15.5$.}
\end{figure}

In Fig.~\ref{f5}
one can see the three point event spectra selected for $\lambda \geq 15.5$ from the spectra presented in Fig.~\ref{f4}. The background in energy region of $(22 \div 29)$ keV for $^{78}$Kr is 15 events for 9457 h$^{-1}$, and for $^{nat}$Kr is 4 events for a period of 6243 h$^{-1}$ (or 6 events for 9457 h$^{-1}$).
Therefore, using work of Gary J. Feldman and Robert D. Cousins
\cite{Feldman99}, for a given background (6 events) and effect + background
(15 events), with C.L. 90\%, we obtain
$N_{eff}  = 9_{ - 5.52}^{ + 5.72}$ for 9547 h $(\mu=3.48\div16.52)$,
and finally we have $N_{eff}=8.33$ yr$^{-1}$ (90\% C.L.) for one year.

The half-life period has been calculated using formula:
\begin{equation}
\label{eq1}
\mathrm{T}_{1/2} = \frac{(ln2) \cdot N \cdot p_3 \cdot \varepsilon_p \cdot
\varepsilon_3 \cdot \alpha_k \cdot k_\lambda } {  N_{eff}},
\end{equation}
where $N=1.08\cdot10^{24}$ is the number of $^{78}$Kr atoms in
the operating volume of the counter, $p_3=0.47$ is a portion of $2K$-captures accompanied by the emission of two-quanta;
$\varepsilon_p=0.809$ is the probability of two \emph{K}-quanta
absorption in the operating volume; $\varepsilon_3=0.5$ is the selection efficiency for three-point events due to $2K$-capture in $^{78}$Kr;
$\alpha_k=0.985$ is the portion of events with two \emph{K}-quanta that
could be registered as distinct three-point events;
$k_\lambda=0.840$ is the useful event selection
coefficient for a given threshold with respect to $\lambda$.
After the substitution of all values into expression (\ref{eq1}) we have:

\vspace{0.2cm}

T$_{1/2}(0\nu+2\nu,2K) = 1.4^{+2.2}_{-0.7}\cdot 10^{22}\texttt{yr}~(90\%~\texttt{C.L.})$ ${}$ ${}$  or \\

T$_{1/2}(0\nu+2\nu,2K) \ge 7\cdot 10^{21}\texttt{yr}~(90\%~\texttt{C.L.})$ \ \ \ \
\vspace{0.2cm}

This work was supported by the Russian Foundation for Basic Research (Grant No. 11-02-00761a) and the Neutrino Physics Program of the Presidium of the Russian Academy of Sciences.


\end{document}